\documentclass[]{spie}  %>>> use for US letter paper
\usepackage{amsmath,amsfonts,amssymb}
\usepackage{graphicx}
\usepackage[colorlinks=true, allcolors=blue]{hyperref}

\title{Reaching sub-milimag photometric precision on Beta Pictoris with a nanosat: the PicSat mission}

\author[a]{M. Nowak}
\author[a]{S. Lacour}
\author[a]{V. Lapeyr\`ere}
\author[a]{L. David}
\author[a]{A. Crouzier}
\author[a]{C. Dufoing}
\author[a]{H. Faiz}
\author[b]{T. Lemoult}
\author[c]{P. Tr\'ebuchet}

\affil[a]{LESIA, Observatoire de Paris, PSL Research University, CNRS, Sorbonne Universités, UPMC Univ. Paris 06, Univ. Paris Diderot, Sorbonne Paris Cité}
\affil[b]{Chelles Observatory, 23 avenue H\'enin, 77500 Chelles France}
\affil[c]{Agence nationale de la s\'ecurit\'e des syst\'emes d'information, 75700 Paris France}
\begin{document} 
\maketitle

\begin{abstract}

PicSat is a nanosatellite currently being developed to observe the transit of the giant planet $\beta$ Pictoris, expected some time between July 2017 and June 2018. The mission is based on a Cubesat architecture, with a small but ambitious 2 kg opto-mechanical payload specifically designed for high precision photometry. The satellite will be launched in early 2017, probably on a 600 km Sun synchronous orbit. The main objective of the mission is the constant monitoring of the brightness of β Pic at an unprecedented combination of reliability and precision (200 ppm per hour, with interruptions of at most 30 minutes) to finely characterize the transiting exoplanet and detect exocomets in the β Pictoris system. To achieve this difficult objective, the payload is designed with a 3.5 cm effective aperture telescope which injects the light in a single-mode optical fiber linked to an avalanche photodioode. A two-axis piezoelectric actuation system, driven by a tailor-made feedback loop control algorithm, is used to lock the fiber on the center of the star in the focal plane. These actuators complement the attitude determination and control system of the satellite to maintain the   sub-arcsecond pointing accuracy required to reach the excellent level of photometric precision. Overall, the mission raises multiple very difficult challenges: high temperature stability of the avalanche detector (achieved with a thermoelectric colling device), high pointing accuracy and stability, and short timeframe for the development.

\end{abstract}


\section{INTRODUCTION}
\label{sec:introduction}

Beta Pictoris is a young ($\sim 12\text{ Myr}$) and close-by (20 pc) star which, in recent years, has gained a considerable notoriety among astronomers. The presence of a circumstellar debris disk, detected in 1984, and of a young giant planet ($\beta$ Pic b, Lagrange et al.\cite{Lagrange2009}), makes this system an ideal target for the study of planet formation.

A number of studies have helped to unravel some of the mysteries of $\beta$ Pictoris. Observations made with VLT/NaCo in 2009 (after the initial discovery in 2003) showed that the planet was a giant (6 to 12 $M_\text{Jup}$) orbitting between 8 to 15 Astronomical Units (AU) from its parent star. These orbital parameter estimates were later refined by Chauvin et al.\cite{Chauvin2012} and then by Millar-Blanchaer et al.\cite{Millar-Blanchaer2015}. The semi major-axis was measured between 8 to 9 AU, and the eccentricity $\le{}0.26$. The mass of the planet was found consistant with ``hot start'' evolutionnary model and with J ($1.265~\mu{}\mathrm{m}$), H($1.66~\mu{}\mathrm{m}$), and M'($4.78~\mu{}\mathrm{m}$) band photometry obtained with VLT/NaCo (Bonnefoy et al.\cite{Bonnefoy2013}). Because all these studies have been made using direct light observations, the diameter of the planet remains unknown.

The discovery of $\beta$ Pictoris was announced in 2009, but its existence has already been suggested by Lecavelier et al.\cite{Lecavelier1995, Lecavelier1997} in the late 1990s. Based on the observation of a photometric event detected in 1981, they suggested that the $\beta$ Pictoris system could host a transitting planet. In recent works, Lecavelier et al.\cite{Lecavelier2009, Lecavelier2016} showed that the orbit of $\beta$ Pic b was indeed consistant with a 90 degree inclination of the orbital plane, and with a time of transit some time in 1981. If this is really the case, then the planet is expected to transit in mid-2017 or early 2018 (depending on whether the planet is orbitting on a low eccentricity orbit, with a period of 18~yr , or on a high eccentricty orbit, with a period of 36 yr).

Given the uncertainty on the time of the transit (one year), and the total duration of such an event \hbox{($\sim{}10\text{ hours}$)}, catching it requires a dedicated year-long monitoring. But $\beta$ Pictoris is only visible from the Southern hemisphere, and is hidden by the Sun in summer (which, according to current predictions, is currently the most likely time for this transit). Consequently, catching the event with a ground-based telescope seems impracticle. And despite the obvious scientific interest of an observation of this transit, which would allow to directly measure the diameter of the planet and give precious information on its close-by environment (Hill sphere), the typical timeframe and development costs of space-based observatories could lead to the conclusion that a dedicated space mission is unrealistic. But the emerging technology of nanosatellites changes that.

During the past decade, the number of small satellites ($\le{}10$ kg) has increased drastically. Initially mainly used by universities as a mean for student to get some hands-on experience on the design and realization of a space mission, they have now found their way to the military and commercial sectors\footnote{https://sites.google.com/a/slu.edu/swartwout/home/cubesat-database}. For laboratories, they offer the possibility of creating small low-cost short-timeframe missions.

The PicSat mission falls into the latter category. The mission is a 3-Unit (3U) CubeSat ($30\text{ cm} \times{} 10\text{ cm} \times{} 10\text{ cm}$, $\sim$ 4 kg), whose goal is the constant monitoring of the visible photometry of $\beta$ Pictoris to detect the transit of the planet b expected in mid-2017/mid-2018. Here we present a general overview of the mission and its current status. In section~\ref{sec:science}, we give the main requirements of the mission, based on its science objectives. Section~\ref{sec:mission} focuses on the design of the satellite and its opto-mechanical payload. In section~\ref{sec:pointing} we give some details on one of the major problems of this mission: pointing and control. Our conclusions are given in section~\ref{sec:conclusion}.

\section{SCIENCE OBJECTIVES}
\label{sec:science}

\subsection*{Transit of the giant planet $\beta$ Pictoris b}

The main science objective of the PicSat mission is the observation and the characterization of the transit of $\beta$ Pictoris b, expected between July 2017 and June 2018. Based on the event detected in 1981 in the light curve of the star, and on current estimates of the orbital path of the planet, the transit is expected to have a total duration of $\sim$ 10 hours, and a contrast of a few percents at most. A 7-$\sigma$ certainty detection of the transit requires to be able to achieve a photometric precision of a few $10^{-3}/\text{hour}$.

This level of precision will be enough to claim the detection of the transit, but if we want to acquire useful science data on the planet, we need better than that. In 1981, the sharp decrease of the flux of $\beta$ Pic, which was interpreted as a transitting planet, was preceeded by a slow and steady rise of the brightness, up to 0.06 magnitude (about 5\% of the flux). It has been suggested that this was a sign of the presence of a gap in the debris disk due to the presence of the planet. Studying this particular pattern at a high temporal resolution could help better characterize the close environment of the planet, especially its Hill sphere, and help understand the interactions between the planet and the disk. To do so, the mission would require a photometric precision at least one order of magnitude better than what it necessary to detect the transit (i.e. $10^{-4}/\text{hour}$).

\subsection*{Transitting exocomets}

A secondary science goal of PicSat is the detection of exocomets in large band transit observations. Exocomets have already been detected in some star systems (see for example the work of Kiefer et al.\cite{Kiefer2014}), but by means of transit spectrometry, where simultaneous absorption features are detected in different lines (in the latter case, the CA~II, K, and H lines). The feasibility of detecting exocomets in large band observations remains to be demonstrated. Lecavelier et al.\cite{Lecavelier1999} have developped a model for transitting exocomets which suggests that a photometric precision of $\sim{}10^{-4}/\text{hour}$ would indeed allow the detection of exocomets with realistic gas production rates.

\subsection*{Disk inhomogeneities}

The last science objective of PicSat is the study of disk inhomogeneities by the constant monitoring of the brightness of $\beta$ Pic. As a by-product of its quest for the expected transit and/or exocomets, PicSat will produce an homogenous and continuous set of photometric measurements of $\beta$ Pic, which will provide precious information on the structure of the debris disk.

\section{THE PICSAT MISSION}
\label{sec:mission}

From the science objectives, we know that our mission has to be able to achieve a photometric precision of $10^{-4}/\text{hour}$, and should constantly observe $\beta$ Pic from July 2017 to June 2018, acquiring at least 1 point every hour or so. Hereafter we present the design of the mission made to achieve these objectives.

\subsection{Orbit}

The orbit of the satellite is highly constrained by Sun exposition (for power generation), observability of $\beta$ Pictoris (\hbox{RA 05h 47min 17s}, \hbox{$\delta= -51\text{deg}~03'~59''$}), ground station visibility, and launch opportunities. Most of the launch opportunities available concern polar Sun-Synchronous Orbit (SSO), very well-suited for Earth observation. In a SSO, the natural precession of the orbital plane induced by the oblateness of the Earth exactly counterbalance the rotation of the Earth around the Sun so that the illumination angle is constant over the year. For our own mission, we studied different possibilities, with altitudes ranging from 500~km to 650~km, and found that in all cases, the total fraction of $\beta$ Pic visibility ($\ge{}63\%$), and Sun exposition ($\ge{}75\%$) was compatible with our objectives. The final choice will depend on launch opportunities for Q2 2017.

\subsection{Payload}

\begin{figure}
  \begin{center}
    \includegraphics[width=0.75\linewidth]{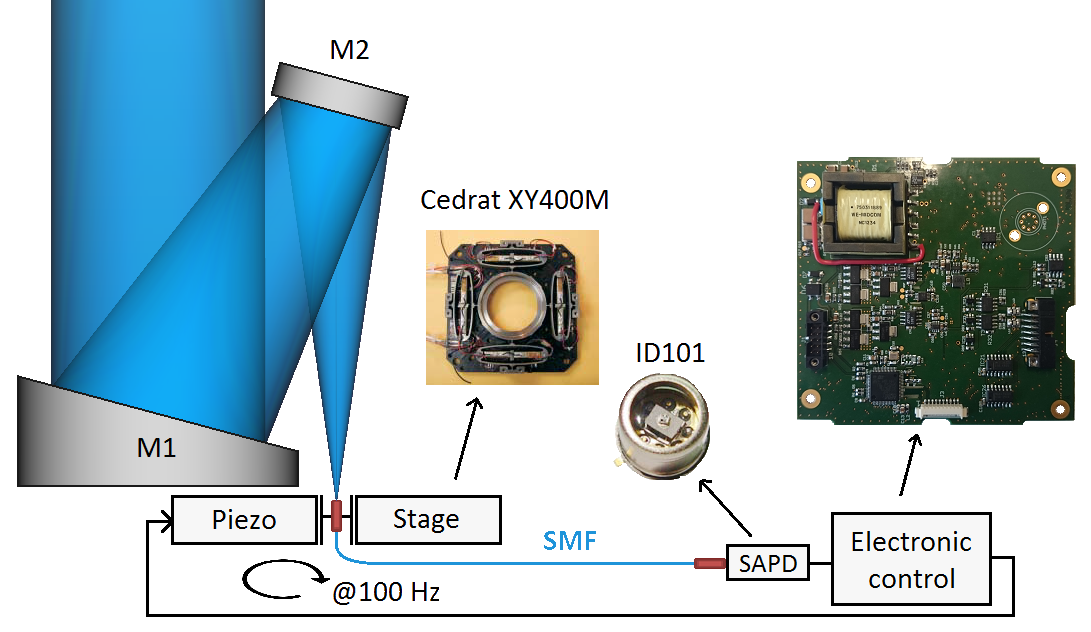}
    \caption{Principle of the PicSat opto-mechanical payload}
    \label{fig:payload}
  \end{center}
\end{figure}

The payload of the PicSat mission is represented in Figure~\ref{fig:payload}. It is designed to achieve high precision photometry ($10^{-4}/\text{hour}$) in the visible band on $\beta$ Pic ($M_V=3.86$). To do so, it uses a 3.5 effective diameter (5 cm real) optical telescope, with a 13.5 cm focal length. The light is collected by a primary off-axis parabola (M1) and reflected onto a secondary plane miror (M2). In the focal plane, the flux is collected by a single-mode optical fiber (SMF) which acts as a very strong spatial filter, rejecting most of the scattered light. The fiber then bring the light to a photon counting Single Pixel Avalanche Photodiode (SAPD) which is temperature-regulated by a thermo-electric module for photometric stability. The fiber head is mounted on a two-stage piezo-electric actuator, which is loop-controled to track the star and keep the injected flux stable (see Section~\ref{sec:pointing}). The entire payload is controled by a specifically designed electronic board. A 72~MHz STM32F303 microchip drives all the different subsystems (including the control loop for the piezo stage, the SAPD and its temperature regulation, and all the house-keeping sensors). The whole mechanical structure is designed in aluminum only, so as to keep the thermal deformation homogenous, and avoid strong variations of the Point Spread Function (PSF) of the instrument. 

The estimated photometric budget is given in Table~\ref{tab:budget}. The system, based on a photon counting detector, is inherently free of read noise. Scattered light is kept at a very low level thanks to the spatial filtering introduced by the single mode optical fiber. The two main sources of noise (excluding the inevitable photon noise) are due to: thermal variations of the SAPD efficiency, and stability of the fine pointing loop (i.e. stability of the fiber injection). 

\begin{table}
  \begin{center}
    \begin{tabular}{l l c}
      \hline
            \hline
      Noise source  & Assumption & Error (ppm/hour) \\
      \hline
      Photon noise  & $M_V=3.86$ & 60 \\      
      Readout noise  & No readout noise & 0 \\
      Sattered light & 150$e^-/s$ & 80 \\
      SAPD bias voltage stability  & $100\mu\text{V}$ & 20 \\
      Thermal stability & Regulated and corrected to $0.01^{\circ}\mathrm{C}$ & 40 \\
      Injection stability & 5\% at 100 Hz & 80 \\
      \hline
            \hline
            & \textbf{Total} & \textbf{140}\\
            & & \\
    \end{tabular}
    \caption{Estimated photometric budget}
    \label{tab:budget}
  \end{center}
\end{table}

\subsection{Satellite Platform}

\begin{figure}
  \begin{center}
  \includegraphics[width=0.8\linewidth]{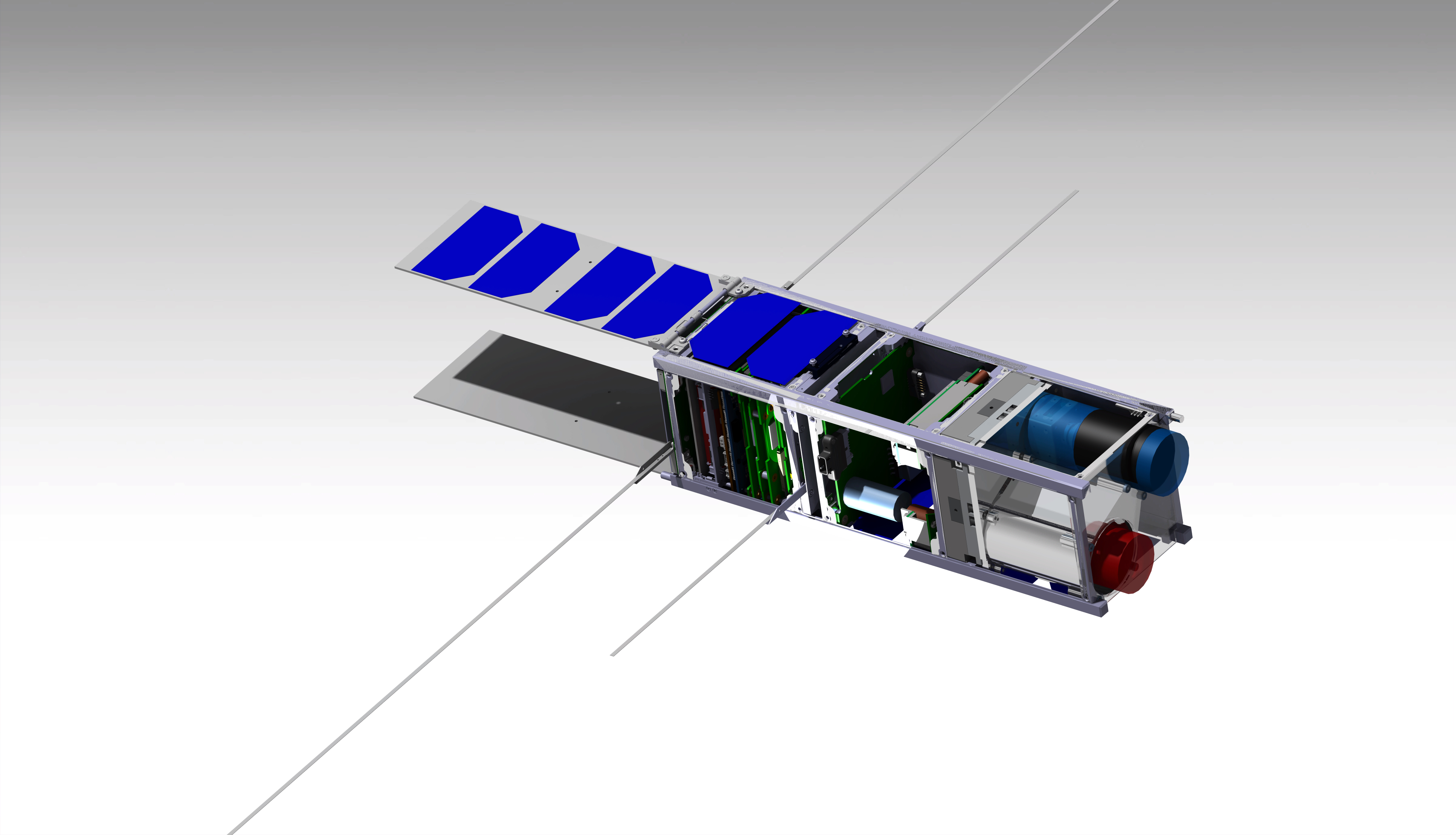}
  \caption{A 3D rendering of the satellite, based on a 3U Cubesat structure.} 
  \label{fig:platform}
  \end{center}
\end{figure}

To support the payload, and provide all the necessary power, data handling, and communication capabilities, a satellite platform has been designed. The general architecture is given in Figure~\ref{fig:platform}. The design is based on a 3U Cubesat structure. The first (right) unit is dedicated to the optical telescope and to the piezo-electric stage. It also hosts the star tracker, which is the main sensor used for attitude control and pointing. The mechanical design ensure a rigid link between the payload and the star tracker, so that even if the structrure bends under thermal stress, the attitude measurement provided by the star tracker will reflect the real pointing of the payload.

The Attitude Determination and Control System (ADCS) is located in the second (middle) unit of the satellite, along with the payload board and the SAPD. The ADCS is a complete, yet very compact, system manufactured by Hyperion Technologies\footnote{www.hyperiontechnologies.nl}. It includes: a 3-axis gyroscope and magnetometer, 3 reaction wheels (one per axis), and 3 magnetotorquers (one per axis). The system will provide a 30 arcsec 1-$\sigma$ pointing precision.

The On-Board Computer is located in the left unit, along with the power, communication, and data handling systems. The OBC is based on an ARM9 400 MHz processor, and will be in charge of running all satellite operations. The OBC will also be responsible for some on-board processing of the data collected. For ground communications, a set of 2 UHF / 2 VHF dipole antennas are used. All these systems are of-the-shelf systems, manufactured by ISIS\footnote{www.isispace.nl}.

\section{POINTING AND CONTROL}
\label{sec:pointing}

As mentionned in Section~\ref{sec:mission}, one of the main source of noise on the photometric flux comes from fluctuations of the fiber-telescope coupling. The pointing precision of the satellite platform is constrained by the ADCS capabilities to $\sim$30 arcsec (1-$\sigma$). In the focal plane, this creates x-y oscillations of the position of the target star which extends to $\sim{}20\mu\text{m}$. For comparison, the PSF Full Width at Half-Maximum (FWHM) is $\sim{}3\mu\text{m}$. Even taking into account the broadening created by the injection in the single-mode fiber, the effective FWHM of the PSF is still $\le{}4\mu{}\text{m}$.

To track the star at a better precision, the fiber is mounted on a 2-axis loop-controled piezo-actuator. The total range accessible is $\pm{}250\mu{}\text{m}$, which corresponds to a field-of-view of about $\pm{}1'$ on the sky. When the satellite points the star, the ADCS ensure that the star is actually in the accessible range of the piezo. The payload board starts a ``scanning'' algorithm to find the location of the target in the focal plane. When the target is acquired, the payload switches to a ``tracking'' mode.

\begin{figure}
  \begin{center}
    \includegraphics[width=0.4\linewidth]{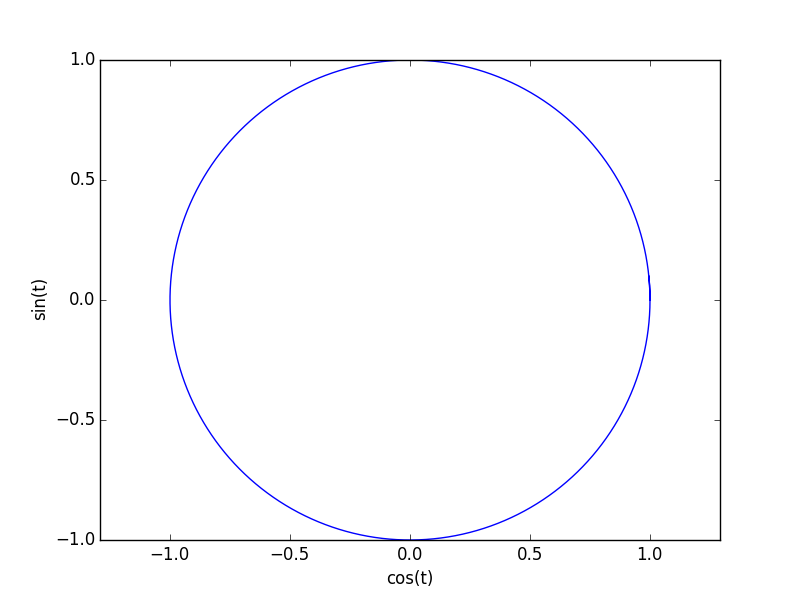}
    \includegraphics[width=0.34\linewidth]{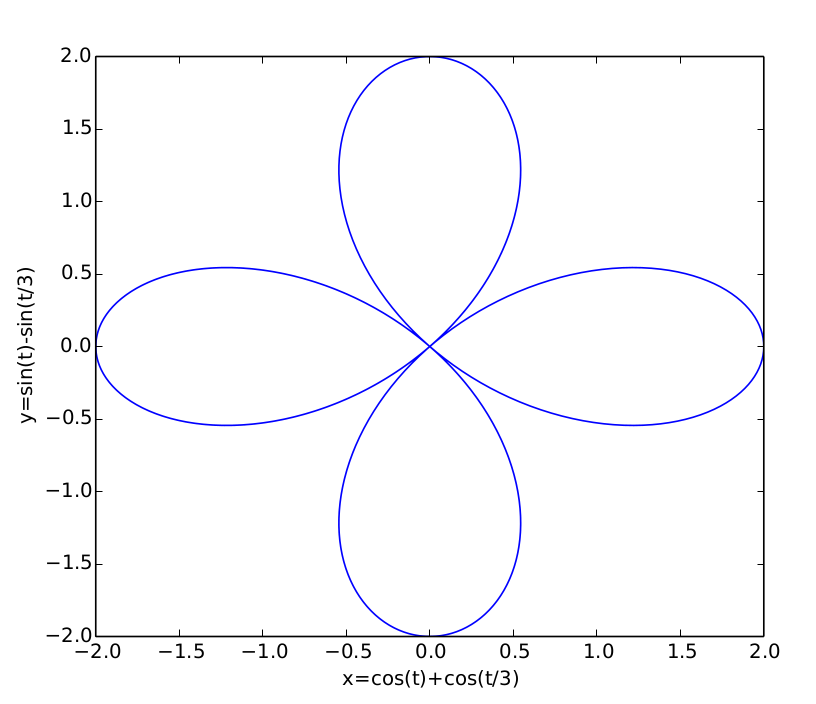}
    \caption{Two of the modultaion patterns currently under study. The first modulation is a simple circular trajectory, which ensure a constant monitoring of the tail of the PSF (and thus of its FWHM in every directions). The second modulation has the shape of a four-leafed clover. It only scans the FWHM in two orthogonal directions, but it also spends more time in the very central region of the PSF (thus giving higher mean flux). In each case, the modutlation is run at 100 Hz.}
    \label{fig:modulation}
  \end{center}
\end{figure}

Because the $10^{-4}/\text{hour}$ photometric objective requires to be able to estimate the possible deformations of the PSF (due mainly to thermal stress in the optics), the ``tracking'' algorithm has actually two distinct objectives: it is used to follow the motion of the star in the focal plane (so that the star is never ``lost''), and it also introduces a ``modulation'' around the estimated central position of the star, so that the shape of the PSF can be estimated during data-processing. The fiber is moved in the focal plane at a frequency of 1000 Hz, and its position is modulated at 100 Hz around the position of the star:
\begin{eqnarray*}
  x(t)=\hat{x}_*(t)+x_\text{mod}(t)\\
  y(t)=\hat{y}_*(t)+y_\text{mod}(t)
 \end{eqnarray*}
\noindent{}where $(\hat{x}_*, \hat{y}_*)$ is the real-time estimation of the position of the star, and $(x_\text{mod}, y_\text{mod})$ is a 100 Hz modulation around this position. Typical modulation schemes currently under study are given in Figure~\ref{fig:modulation}.

Two different algorithms are currently being studied for this mode. A first algorithm uses a combination of measurements made with the SAPD (the flux measured in real time gives some information on the distance of the star relative to the fiber core), and of measurements made with a 3-axis gyroscope embedded on the payload board. The sensor fusion is made through an extended Kalman filter running at 1~kHz.

In the second algorithm, the 3-axis gyroscope is not used. Because this sensor has a rather low accuracy (0.01 degree/s), and is located in the second unit of the Cubesat (on the payload board, see Section~\ref{sec:mission}), there is a non-negligible possibility that its measurements will be unusable when in flight. In such case, another algorithm will be used, based only on the flux measurements. This algorithm will run at 100 Hz, and will estimate the position of the star relative to the fiber using only flux measurements. To do so, a set of 10 values of the flux along the modulation path will first be acquired. Then a least-square estimator will be used to find the position that best fits these measurements. This will give an estimate of the central position of the star that will then be used as the central position of the next modulation pattern.

\begin{figure}
  \begin{center}
    \includegraphics[width=0.45\linewidth, trim=17cm 0cm 17cm 0cm, clip]{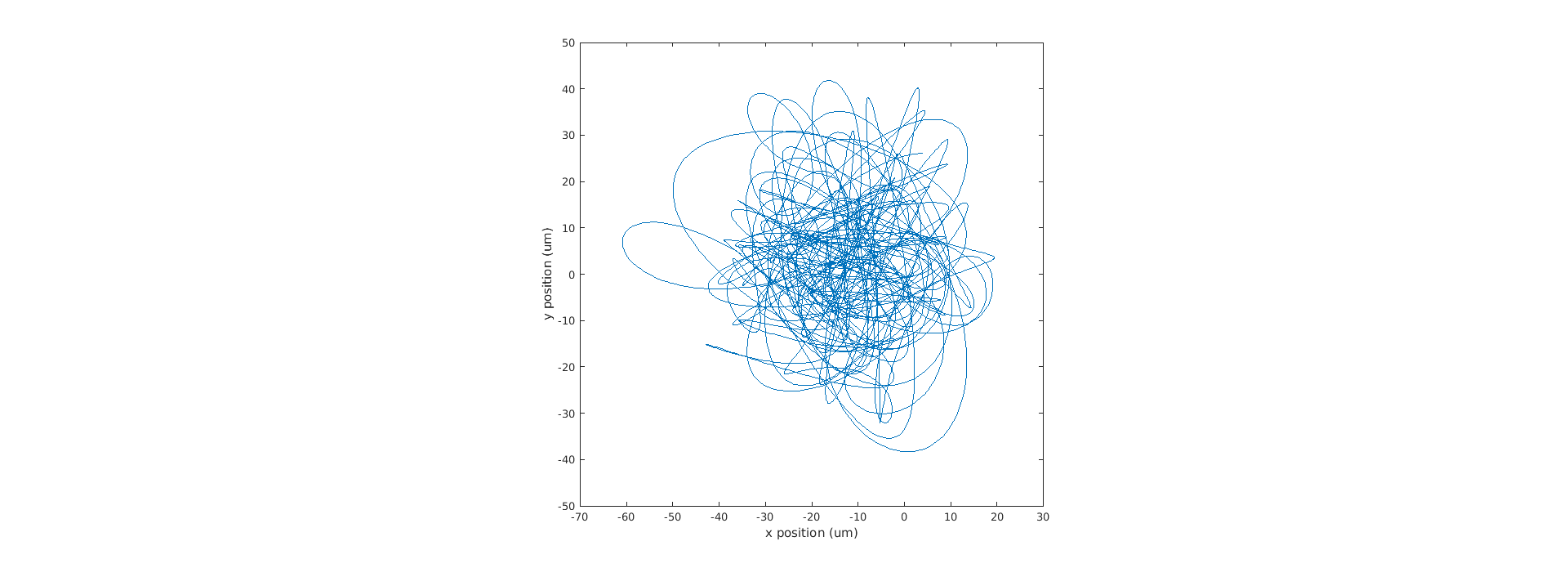}
    \includegraphics[width=0.45\linewidth, trim=17cm 0cm 17cm 0cm, clip]{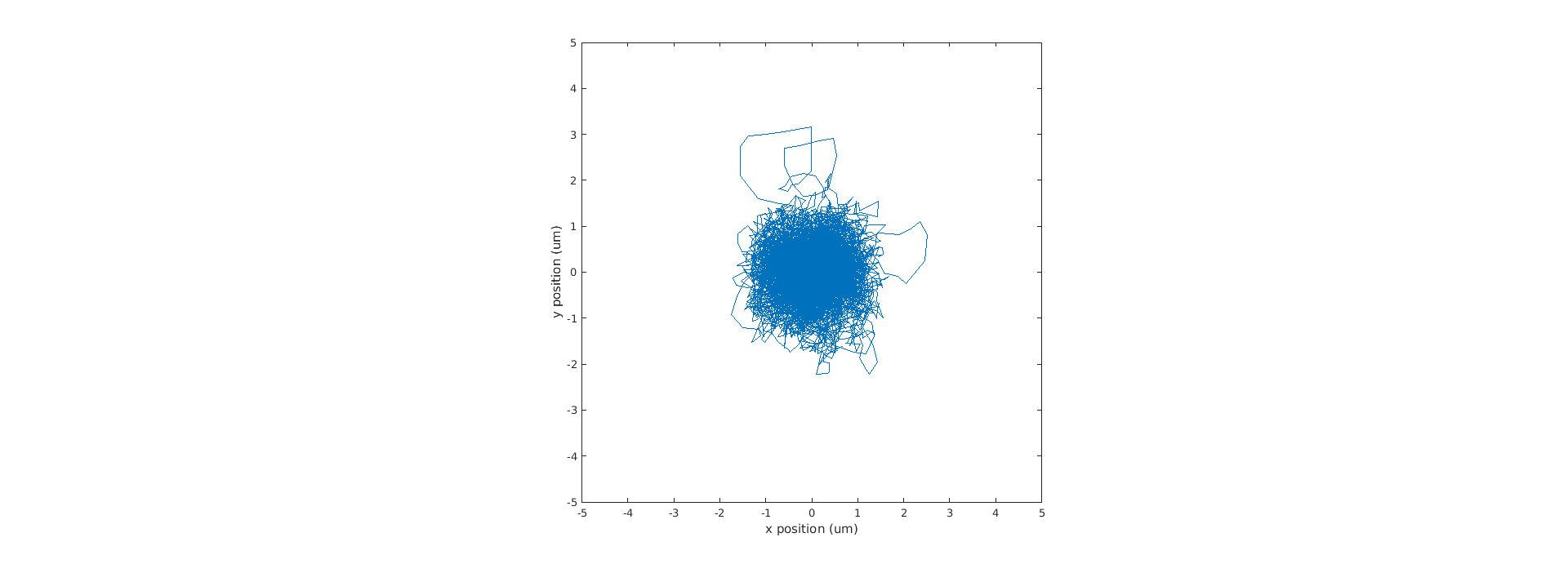}
    \caption{Displacement of the star relative to the fiber with (left) and without (right) the loop-controled piezo-actuators. In the left panel, the model used for the displacement of the star is a white noise with frequencies ranging from 0 to 10 Hz, and with a 1-$\sigma$ excursion of $20\mu{}\text{m}$. In the right panel, the measured 1-$\sigma$ value is $0.7\mu{}\text{m}$ on both axes.}
    \label{fig:simu}
  \end{center}
\end{figure}

Using a simple model for the noise created by the reaction wheels, we used MATLAB/Simulink to simulate the behavior of the control loop over 10 s, and found that even in the presence of photon noise, the total 1-$\sigma$ displacement of the star could be brought down to $\le{}1\mu{}\text{m}$ (i.e. $\sim$~1~arcsec), as shown in Figure~\ref{fig:simu}. Our model still lacks a realistic transfer function for the piezo actuators, as well as a model accouting for thermal low-frequency deformations of the PSF, but the results obtained so far are very encouraging.

\section{CONCLUSION: CURRENT STATUS OF THE MISSION}
\label{sec:conclusion}

Here we gave a general overview of the design of our 3U Cubesat dedicated to the observation of the transit of $\beta$ Pictoris b, expected in mid-2017/mid-2018.

A taylor-made opto-mechanical payload has been designed to achieve a $10^{-4}/\text{hour}$ photometric precision on the visible flux of $\beta$ Pic. This payload uses a 3.5 cm effective diameter optical telescope which injects the light into a single mode optical fiber linked to an avalanche photodiode. The system is free of reading noise, and thanks to the single mode fiber, scattered ligth is largely filtered out. Thus, along with quantum photon noise, the main source of photometric noise comes from fuctuations of the injection in the fiber (i.e. pointing precision and stability).

Preliminary tests have shown that a 1 arcsecond pointing precision was not out-of-reach using a two stage system, made of an off-the-shelf ADCS system and a specifically designed loop-controled piezo actuator. A first version of our payload board has been tested in vibrations and thermal vacuum. Radiation tests are planned in the coming months, and a second version of the board is currently being manufactured. We plan on tesing the pointing algorithm on an air-bearing test bench currently under development to demonstrate that the required photometric precision can be achieved. Assembly, integration and test activities of the flight model will start in late 2016, and the satellite could be ready for launch in early 2017. On-orbit testing is scheduled for May/June 2017, and science operations should start in July 2017.

\acknowledgments{}
This  work  was  supported  by the  European  Research Council (ERC-STG-639248), the Fondation MERAC (Mobilising European reserach in Astrophysics and Cosmology), and the Labex ESEP (Exploration Spatiale des Environnements Planétaires).

\bibliographystyle{spiebib} % makes bibtex use spiebib.bst
\bibliography{bibliographie} % bibliography data in report.bib

\end{document}